\documentclass[12pt]{article}
\usepackage{graphicx}
\begin{document}
\noindent {\footnotesize\it  %%ISSN 1063-7737,
 Astronomy Letters, 2009, Vol. 35, No. 12.
 }

\noindent
\begin{tabular}{llllllllllllllllllllllllllllllllllllllllllllll}
 & & & & & & & & & & & & & & & & & & & & & & & & & & & & & & & & & & & & &  \\\hline\hline
\end{tabular}

 \vskip 1.5cm
 \centerline {\large\bf Kinematics of Tycho-2 Red Giant Clump Stars}
 \bigskip
 \centerline {V.V. Bobylev, A.S. Stepanishchev, A.T. Bajkova, and G.A. Gontcharov}
 \bigskip
 \centerline {\small\it
 Pulkovo Astronomical Observatory, Russian Academy of Sciences, St-Petersburg}
 \bigskip

{\bf Abstract--}Based on the Ogorodnikov-Milne model, we analyze
the proper motions of 95 633 red giant clump(RGC) stars from the
Tycho-2 Catalogue. The following Oort constants have been found:
$A = 15.9\pm0.2$ km s$^{-1}$ kpc$^{-1}$ and $B = -12.0\pm0.2$ km
s$^{-1}$ kpc$^{-1}$. Using 3632 RGC stars with known proper
motions, radial velocities, and photometric distances, we show
that, apart from the star centroid velocity components relative to
the Sun, only the model parameters that describe the stellar
motions in the $XY$ plane differ significantly from zero. We have
studied the contraction (a negative $K$ effect) of the system of
RGC stars as a function of their heliocentric distance and
elevation above the Galactic plane. For a sample of distant
(500--1000 pc) RGC stars located near the Galactic plane
($|Z|<200$ pc) with an average distance of $d=0.7$ kpc, the
contraction velocity is shown to be $Kd= -3.5\pm0.9$ km s$^{-1}$;
a noticeable vertex deviation, $l_{xy} = 9^\circ\pm0.5^\circ$, is
also observed for them. For stars located well above the Galactic
plane ($|Z|\geq200$ pc), these effects are less pronounced, $Kd =
-1.7\pm0.5$ km s$^{-1}$ and $l_{xy} = 4^\circ.9\pm0^\circ.6$.
Using RGC stars, we have found a rotation around the Galactic $X$
axis directed toward the Galactic center with an angular velocity
of $-2.5\pm0.3$ km s$^{-1}$ kpc$^{-1}$, which we associate with
the warp of the Galactic stellar-gaseous disk.

\section*{INTRODUCTION}

The red giant clump (RGC) stars are distributed fairly uniformly
over the celestial sphere. Since they occupy a compact region on
the Hertzsprung-Russell diagram and, hence, have the same
luminosity, reliable photometric distance estimates can be
obtained for them. These stars are of great interest in analyzing
the three-dimensional kinematics of various Galactic subsystems.

First of all, we are interested in such peculiarities as the
expansion/contraction of stellar groups, the deviation of their
vertex from the direction of the kinematic center, i.e., the
Galactic center, and the connection of the stellar kinematics with
the warp of the Galactic stellar-gaseous disk.

Various authors have found a negative value of the Oort parameter
K responsible for the expansion/contacrtion. This means that the
system of stars is in the state of contraction. We are talking
about the plane effect that describes the contraction in the
Galactic $XY$ plane. The negative K effect in the motion of
distant OB stars was detected by Torra et al. (2000) and Bobylev
(2004). It was also revealed in the motion of open star clusters
(Bobylev et al. 2007) and in the motion of dwarf stars (Rybka
2004a) and giant stars (Rybka 2004b, 2008).

At present, the nature of the negative K effect is not completely
clear. It is hypothesized that this effect is related to: (i) the
peculiarities of stellar radial velocity measurements (Pont et al.
1994), (ii) the influence of the bar at the Galactic center (Point
et al. 1994), and (iii) the influence of the Galactic spiral
structure (Rohlfs 1977; Metzger et al. 1998). In particular,
Fern\'andez et al. (2001) and Bobylev et al. (2006a) made an
attempt to take into account the influence of a spiral density
wave; it failed, because the negative K effect was retained.

Rybka (2008) showed that the RGC stars have different vertex
deviations and differ significantly by the magnitude of the K
effect, depending on the Galactic latitude. In this respect, it is
interesting to trace the changes in kinematic parameters of these
stars located in various layers parallel to the Galactic plane.

Analysis of the large-scale structure of neutral hydrogen has
shown that the gaseous disk in the Galaxy is warped (for a review,
see Burton, 1988). The results of studying this structure on the
basis of currently available data on the HI distribution are
presented in Kalberla and Dedes (2008). This structure is revealed
by the spatial distribution of stars and dust (Drimmel and Spergel
2001) and by the distribution of pulsars in the Galaxy (Yusifov
2004). Having analyzed nearby Hippacros (1997) stars, Dehnen
(1998) showed that the distribution of their residual velocities
$V_y-V_z$ was in satisfactory agreement with various rotation
models of the warped disk. Miyamoto et al. (1993) and Miyamoto and
Zhu (1998) determined the rotation parameters of the warped
stellar-gaseous disk by analyzing giant stars of various spectral
types and samples of Hipparcos O-B5 stars. Thus, there is positive
experience in solving this problem using data on stars relatively
close to the Sun.

The goal of this paper is to study the kinematic peculiarities of
a wide solar neighborhood on the basis of data on the RGC stars
from the Tycho-2 Catalogue (Hog et al. 2000) selected by
Gontcharov (2008) and, in particular, to establish the connection
of the kinematics of RGC stars with the warp of the Galactic
stellar-gaseous disk.

\section{DATA}

At present, two lists of candidates for RGC stars compiled from a
combination of Tycho-2 and 2MASS data (Skrutskie et al. 2006) have
been published. These are the catalogs by Rybka (2006, 2007) and
Gontcharov (2008).

The catalog by Rybka (2007) contains about 60000 RGC candidates
from Tycho-2 with photometric characteristics from 2MASS with
distances $>100$ pc. The catalog by Gontcharov (2008) contains
about 97000 stars with data from the same catalogs as those of
Rybka but without any distance constraint. Therefore, the number
of stars is larger. In contrast to the catalog by Rybka (2007),
the catalog by Gontcharov (2008) gives a photometric distance
estimate for each of the stars corrected for interstellar
extinction. In this respect, the catalog by Gontcharov (2008) is
more convenient for solving the kinematic problems considered
here.

The radial velocities taken from the PCRV (Gontcharov 2006;
Bobylev et al. 2006b) and RAVE (Steinmetz et al. 2006) catalogs
are given for 4163 stars. These are relatively close stars with
known Hipparcos trigonometric parallaxes.

Figure~1 gives an idea of the three-dimensional spatial
distribution of stars. The left, middle, and right columns present
the distributions of stars in the $XY,XZ,$ and $YZ$ planes,
respectively. The rows from top to bottom provide the
distributions for the complete sample of stars and the samples of
stars with distances $d<400$ pc, $400<d<600$ pc, and $600<d<1600$
pc. The distance ranges were chosen in such a way that each sample
contained approximately the same number of stars. The figures
reveal some structural peculiarities of the samples. Thus, for
example, clumps inclined by $\approx 20^\circ$ to the $X$ axis,
typical of young stars associated with the Gould Belt, are clearly
traceable in the $ZX$ plane (c) and (d).

\section{THE METHODS OF ANALYSIS}

In this paper, we use a rectangular Galactic coordinate system
with the axes directed away from the observer toward the Galactic
center ($l=0^\circ, b=0^\circ$, the $X$ axis or axis 1), along the
Galactic rotation ($l=90^\circ, b=0^\circ$, the $Y$ axis or axis
2), and toward the North Galactic Pole ($b=90^\circ$, the $Z$ axis
or axis 3).

\subsection{The Ogorodnikov-Milne Model}

{\bf Using the three-dimensional velocity field.}
 In the linear Ogorodnikov-Milne model, we adhere to the notation that was used
by Clube (1972, 1973). The observed velocity ${\bf V}(r)$ of a
star with a heliocentric radius vector ${\bf r}$ is described, to
the terms of the first order of smallness $r/R_0\ll 1$, by the
equation in vector form
$$
\displaylines{ \hfill
 {\bf V}(r)={\bf V}_\odot+M{\bf r}+{\bf V'},\hfill\llap(1)\cr}
$$
where ${\bf V}_\odot(X_\odot,Y_\odot,Z_\odot)$ is the velocity of
the Sun relative to the centroid of the stars under consideration
and {$\bf V'$} is the residual velocity of the star. Here, the
residual stellar velocities are assumed to have a random
distribution. $M$ is the displacement matrix whose components are
the partial derivatives of the velocity ${\bf u}(u_1,u_2,u_3)$
with respect to the distance ${\bf r}(r_1, r_2, r_3)$, where ${\bf
u}={\bf V}(R)-{\bf V}(R_0)$, while $R$ and $R_0$ are the
Galactocentric distances of the star and the Sun, respectively.
Then,
$$
\displaylines{\hfill M_{pq}={\left(\frac{\partial u_p} {\partial
r_q}\right)}_\circ, \quad (p,q=1,2,3). \hfill\llap(2)\cr }
$$
All nine components of the displacement matrix can be determined
if the distances, radial velocities, and proper motions of the
sample stars are known. In this case, the conditional equations
are
$$
\displaylines{\hfill
 V_r= -X_\odot \cos b\cos l-\hfill\llap{(3)} \cr\hfill
 -Y_\odot\cos b\sin l-Z_\odot\sin b+
\hfill\cr\hfill
 +r (\cos^2 b\cos^2 l M_{11}
 +\cos^2 b\cos l \sin l M_{12}+
\hfill\cr\hfill
 +\cos b\sin b \cos l  M_{13}
 +\cos^2 b\sin l\cos l M_{21}+
\hfill\cr\hfill
 +\cos^2 b\sin^2 l   M_{22}
 +\cos b\sin b\sin l M_{23}+
\hfill\cr\hfill
 +\sin b\cos b\cos l M_{31}
 +\cos b\sin b\sin l M_{32}+
\hfill\cr\hfill
 +\sin^2 b  M_{33}),
\hfill\cr\hfill
 4.74 r \mu_l\cos b= X_\odot\sin l-Y_\odot\cos l+ ~~~\hfill\llap{(4)} \cr\hfill
 +r (-\cos b\cos l\sin l  M_{11}
 -\cos b\sin^2 l  M_{12}-
\hfill\cr\hfill
 -\sin b \sin l  M_{13}
 +\cos b\cos^2 l M_{21}+
\hfill\cr\hfill
 +\cos b\sin l\cos l  M_{22}+
\hfill\cr\hfill
 +\sin b\cos l  M_{23}),
\hfill\cr\hfill
4.74 r \mu_b=X_\odot\cos l\sin b+ \hfill\llap{(5)} \cr\hfill
 +Y_\odot\sin l\sin b-Z_\odot\cos b+
\hfill\cr\hfill
 +r (-\sin b\cos b\cos^2 l M_{11}-
\hfill\cr\hfill
 -\sin b\cos b\sin l \cos l M_{12}-
\hfill\cr\hfill
 -\sin^2 b \cos l  M_{13}
 -\sin b\cos b\sin l\cos l M_{21}-
\hfill\cr\hfill
 -\sin b\cos b\sin^2 l  M_{22}
 -\sin^2 b\sin l  M_{23}+
\hfill\cr\hfill
 +\cos^2 b\cos l M_{31}
 +\cos^2 b\sin l M_{32}+
\hfill\cr\hfill
 +\sin b\cos b  M_{33}),
\hfill }
$$
where the stellar proper motion components are in mas yr$^{-1}$
(milliseconds per year), the radial velocity $V_r$ is in km
s$^{-1}$, and the heliocentric distance to the star $r$ is in kpc;
we either calculate the latter based on the known trigonometric
parallax $r = 1/\pi$ or take the photometric distance $d_{phot}$.
Equations (3)--(5) contain twelve sought-for unknowns -- three
components of the velocity
 ${\bf V}_\odot(X_\odot,Y_\odot,Z_\odot)$ in km s$^{-1}$ and nine
components of $M_{pq}$ in km s$^{-1}$ kpc$^{-1}$. The system of
equations (3)--(5) is solved by the least squares method.

The Oort constants can be found as follows:
$A=0.5(M_{12}+M_{21})$,
 $B=0.5(M_{21}-M_{12})$,
 $C=0.5(M_{11}-M_{22})$ and
 $K=0.5(M_{11}+M_{22})$; we determine the vertex deviation
 $l_{xy}$ from the relation tan $\tan 2 l_{xy}=-C/A$.
The angular velocity of Galactic rotation can be found from the
relations $(\Omega_Z)_{R_0}=B-A=-M_{12}$. The equations are
written in such a form that the negative sign of the angular
velocity corresponds to the direction of Galactic rotation.

The matrix (tensor) $M$ can be divided into symmetric, $M^+$, and
antisymmetric, $M^-$, parts. Following Ogorodnikov (1965), we call
them the local deformation tensor and the local rotation tensor,
respectively:
$$
\displaylines{\hfill
 M_{\scriptstyle pq}^{\scriptscriptstyle+}=
   {1\over 2}\left( \frac{\partial u_{p}}{\partial r_{q}}+
   \frac{\partial u_{q}}{\partial r_{p}}\right)_\circ,        \hfill\llap(6)\cr\hfill
M_{\scriptstyle pq}^{\scriptscriptstyle-}=
 {1\over 2}\left(\frac{\partial u_{p}}{\partial r_{q}}-
   \frac{\partial u_{q}}{\partial r_{p}}\right)_\circ, \hfill \cr\hfill(p,q=1,2,3).\hfill }
$$

{\bf Using only the proper motions.}
 When only the stellar proper
motions are used, one of the diagonal terms of the local
deformation tensor is known to become uncertain. Therefore, we can
determine only the differences, for example,
$(M_{\scriptscriptstyle11}^{\scriptscriptstyle+}-
 M_{\scriptscriptstyle22}^{\scriptscriptstyle+})$ and
$(M_{\scriptscriptstyle33}^{\scriptscriptstyle+}-
 M_{\scriptscriptstyle22}^{\scriptscriptstyle+})$.
In this case, we use the conditional equations
$$\displaylines{\hfill
4.74 r \mu_{l}\cos b=
       X_{\odot}\sin l-Y_{\odot}\cos l-\hfill\llap(7)
\cr\hfill
 -r(M_{\scriptscriptstyle32}^{\scriptscriptstyle-}\cos l\sin b
   -M_{\scriptscriptstyle13}^{\scriptscriptstyle-}\sin l\sin b
   +M_{\scriptscriptstyle21}^{\scriptscriptstyle-}\cos b+
\hfill\cr\hfill
   +M_{\scriptscriptstyle12}^{\scriptscriptstyle+}\cos 2l\cos b
   -M_{\scriptscriptstyle13}^{\scriptscriptstyle+}\sin l\sin b+
\hfill\cr\hfill
   +M_{\scriptscriptstyle23}^{\scriptscriptstyle+}\cos l\sin b
  -0.5(M_{\scriptscriptstyle11}^{\scriptscriptstyle+}
  -M_{\scriptscriptstyle22}^{\scriptscriptstyle+})\sin 2l\cos b),
\hfill \cr\hfill
4.74 r \mu_b=
    X_{\odot}\cos l\sin b+Y_{\odot}\sin l\sin b-\hfill\llap(8)
\cr\hfill
    -Z_{\odot}\cos b
 +r(M_{\scriptscriptstyle32}^{\scriptscriptstyle-}\sin l
   -M_{\scriptscriptstyle13}^{\scriptscriptstyle-}\cos l-
\hfill\cr\hfill
-0.5M_{\scriptscriptstyle12}^{\scriptscriptstyle+}\sin 2l\sin 2b
   +M_{\scriptscriptstyle13}^{\scriptscriptstyle+}\cos l\cos 2b+
\hfill\cr\hfill
   +M_{\scriptscriptstyle23}^{\scriptscriptstyle+}\sin l\cos 2b
-0.5(M_{\scriptscriptstyle11}^{\scriptscriptstyle+}
    -M_{\scriptscriptstyle22}^{\scriptscriptstyle+})\cos^2 l\sin 2b+
\hfill\cr\hfill
+0.5(M_{\scriptscriptstyle33}^{\scriptscriptstyle+}
    -M_{\scriptscriptstyle22}^{\scriptscriptstyle+})\sin 2b).
\hfill }
$$
$M_{\scriptscriptstyle21}^{\scriptscriptstyle-}$ is equivalent to
the Oort constant $B$. In accordance with the chosen coordinate
system, the positive directions of rotation are: from axis 1 to 2,
from axis 2 to 3, and from axis 3 to 1.
$M_{\scriptscriptstyle12}^{\scriptscriptstyle+}$ is equivalent to
the Oort constant A. The diagonal components of the local
deformation tensor
$M_{\scriptscriptstyle11}^{\scriptscriptstyle+},
  M_{\scriptscriptstyle22}^{\scriptscriptstyle+},
  M_{\scriptscriptstyle33}^{\scriptscriptstyle+}$ (by definition, they coincide
with $M_{11},M_{22},M_{33}$) describe the general contraction or
expansion (depending on the sign) of the entire stellar system.
Thus, the system of conditional equations (7)--(8) includes eleven
sought-for unknowns to be determined by the least squares method.

Although the unknowns
$M_{\scriptscriptstyle13}^{\scriptscriptstyle-}$ and
$M_{\scriptscriptstyle13}^{\scriptscriptstyle+}$ as well as
$M_{\scriptscriptstyle32}^{\scriptscriptstyle-}$ and
$M_{\scriptscriptstyle23}^{\scriptscriptstyle+}$ in Eq. (7) being
determined cannot be separated between themselves, because they
have the same coefficients ($\sin l\sin b$ and $\cos l\sin b$,
respectively), all variables can be separated through the
simultaneous solution of the system of equations (7) and (8).

The Oort constant
 $K=(M_{\scriptscriptstyle11}^{\scriptscriptstyle+}+
     M_{\scriptscriptstyle22}^{\scriptscriptstyle+})/2$
can be calculated on the basis of the quantities found by solving
the system of equations (7)--(8) as follows:
$$\displaylines{\hfill
 K=((M_{\scriptscriptstyle11}^{\scriptscriptstyle+}-
      M_{\scriptscriptstyle22}^{\scriptscriptstyle+})-
    2(M_{\scriptscriptstyle33}^{\scriptscriptstyle+}-
      M_{\scriptscriptstyle22}^{\scriptscriptstyle+}))/2\hfill\llap(9)}
$$
by assuming $M_{\scriptscriptstyle33}^{\scriptscriptstyle+}$ to be
close to zero. Therefore, we first determine
$M_{\scriptscriptstyle33}^{\scriptscriptstyle+}$ by solving the
system of equations (3)--(5) based on the sample of RGC stars with
known trigonometric parallaxes, radial velocities, and proper
motions to test this assumption. We determine the vertex deviation
$l_{xy}$ from the relation
 $\tan 2 l_{xy}=-C/A=-0.5(M_{\scriptscriptstyle11}^{\scriptscriptstyle+}-
  M_{\scriptscriptstyle22}^{\scriptscriptstyle+})/M_{\scriptscriptstyle12}^{\scriptscriptstyle+}$.

\subsection{The Statistical Method}

We use the well-known statistical method (Trumpler and Weaver
1953; Parenago 1951, 1954) to determine the parameters of the
residual velocity (Schwarzschild) ellipsoid. It consists in
determining the symmetric tensor of moments or the tensor of
residual stellar velocity dispersions. When simultaneously using
the stellar radial velocities and proper motions to find the six
unknown components of the dispersion tensor, we have six equations
for each star. The semiaxes of the residual velocity ellipsoid,
which we denote by $\sigma_{1,2,3}$ can be determined by analyzing
the eigenvalues of the dispersion tensor. We denote the directions
of the principal axes of this ellipsoid by $l_{1,2,3}$ and
$b_{1,2,3}$.

\section{RESULTS}

The results of solving the system of equations (3)--(5) that we
obtained based on the sample of 3632 RGC stars with known
trigonometric parallaxes, radial velocities, and proper motions
are presented in Table 1. The kinematic parameters were found
using both trigonometric parallaxes and photometric distances with
the constraints $e_\pi/\pi<1$ and $\pi>1$ mas. A preliminary
analysis shows that we need these constraints to eliminate the
stars that spoil the statistics and affect significantly the
parameters being determined.

As we see from Table 1, $M_{33}$ is found to be statistically
insignificant. Therefore, we have every reason to neglect $M_{33}$
below when determining the K effect from Eq. (9) based only on the
stellar proper motions.

The parameters of the Ogorodnikov-Milne model obtained on the
basis of 95633 RGC stars are presented in Table 2. Based on the
results of Gontcharov (2008) and the data of Fig. 1, we may
conclude that the spatial distribution of very distant RGC stars
is highly nonuniform. Using RGC stars farther than 1 kpc can lead
to undesirable selection effects. Moreover, as was shown by Olling
and Dehnen (2003), the nonuniformity of the distribution in
Galactic longitude affects noticeably the determination of
kinematic parameters, in particular, the Oort constants.
Therefore, we consider stars closer than 1 kpc. Columns nos. 1--3
in Table 2 give the parameters obtained with various constraints
on the heliocentric distance of a star $d(d_{phot})$; columns nos.
4 and 5 give the parameters for stars with distances of 500--1000
pc located near the Galactic plane ($|Z|<200$ pc) and higher
($|Z|\geq200$ pc), respectively.

We determined the parameters of the residual velocity ellipsoid
based on the statistical method both for all stars of the catalog
and for individual samples. The first column in Table 3 gives the
parameters of the residual velocity ellipsoid for the sample of
3632 stars considered above. We estimated the errors in the
parameters obtained from this sample based on the well-known
method by Parenago (1951) using the fourth-order moments.

The complete sample of stars was divided into individual groups by
the physical properties of the stars themselves as well as by the
heliocentric distance d and the elevation above the Galactic plane
$|Z|$. The results are presented in columns nos. 2--4 of Table 3
in the former case and in Table 4 and Fig. 2, which shows the
residual velocity dispersions as a function of $|Z|$ found by the
statistical method using the proper motions of 95633 RGC stars, in
the latter case.

We divided the RGC stars by their physical properties based on the
selection parameters ``Sel1'' and ``Sel2'' listed in the last
columns of the catalog by Gontcharov (2008). The first parameter
points to the displacement of a star from the center of the
distribution of young RGC stars on the $(B-V)_0$--reduced proper
motion diagram, while the second parameter points to the
displacement of a star from a similar diagram for old RGC stars
(Gontcharov 2008). We formed three groups: group (1) consists of
stars with a difference $0<\Delta S,$ where $\Delta S$ =
Sel1-Sel2, which includes old normal-metallicity stars; the stars
of group (2) have $-0.38<\Delta S<0$ and belong to old
low-metallicity stars; finally, group (3) with $-1.0 <\Delta S <
-0.38$ contains young normal metallicity stars. Consequently, the
third and second groups include the youngest and oldest stars,
respectively. The increase in average $|Z|$ as we pass from the
relatively young groups to the older ones:
 $\overline{|Z|}_{(3)}=183$ pc, $\overline{|Z|}_{(1)}=201$ pc, $\overline{|Z|}_{(2)}=221$ pc,
indicates that this division identifies the stars by age
statistically correctly. Since only the proper motions of RGC
stars were used in this case (their space velocities $U,V,W$
cannot be determined), we estimated the errors in the parameters
of the residual velocity ellipsoid given in column no. 5 of Table
3 statistically, based on the Monte Carlo method.

Figure 3 shows the distributions of residual UV velocities
smoothed by the method described by Bobylev et al. (2006b) for
groups (1), (2), and (3) (separated by the parameter ?S); we used
only 3632 RGC stars with known trigonometric parallaxes, radial
velocities, and proper motions (see Table 1). As we see from the
figures, in comparison with sample (3) of ``young'' stars, the UV
distributions for samples (1) and (2) of ``old'' stars exhibit a
larger velocity dispersion manifested as the ``spreading'' of
peaks and the appearance of new clumps, which is caused by the
prolonged action of some dynamical factor (a bar, a spiral density
wave, and the like). However, it should be noted that the velocity
dispersions calculated using the known trigonometric parallaxes
rather than the photometric distances (as with the complete
sample) show the highest values for sample (1), not for (2), which
is in conflict with the results of Table. 3. Qualitatively, the
older age of the stars from group (1) follows from a comparison of
the $UV$-velocity distributions shown in Fig. 3.

Consequently, the method of selection by the parameter $\Delta S$
we used is a reliable method for identifying only the
comparatively ``young'' fraction of stars. A reliable separation
of stars with various ages among the remaining ``old'' fraction of
stars is unlikely to be possible, as suggested by the
insignificant difference in velocity dispersions for groups (1)
and (2), while the velocity dispersions for the young fraction of
stars differ significantly. In addition, we obtained contradictory
results of the selection by the parameter $\Delta S$ for the
entire sample of $\approx$97 000 stars and the sample of 3632
stars with known parallaxes, radial velocities, and proper
motions. Clearly, groups (1) and (2) are actually a mixture of
stars with various ages.

In Fig. 4, the contraction velocity $Kd$ is plotted against the
heliocentric distance $d$.

The parameters being determined depend on the division of the
original sample into subsamples according to the measured
distances. Since all distances contain errors, on average, they
show systematic deviations from the true distances. These
deviations correlate with the measured distances themselves (the
largest measured distances are, on average, overestimated, while
the smallest ones are, on average, underestimated).

Therefore, to ascertain the dependence of the kinematic parameters
being determined on the distances including the measurement
errors, we performed statistical Monte Carlo simulations. To this
end, we solved the system of equations (7) and (8) with
artificially introduced random errors in the distances distributed
normally. The division into subsamples was made by the derived
model distances. We analyzed all stars of the catalog without any
constraints, but the star was rejected in the case where a
negative distance emerged. The equations were solved for 100
realizations for each subsample. We considered three cases: (i)
without introducing any errors, so that these results are a good
addition to the data of Table 2; (ii) the model errors are 20\% of
the measured distance, which is a typical error in the photometric
distance; (iii) the model errors are 60\% of the measured distance
(considered as the extreme case).

The simulation results are presented in Figs. 5, 6, and 7, where
all eleven parameters of the Ogorodnikov-Milne model are plotted
against the distance. Figures 5a, 6a, and 7a show the velocity
components of the Sun relative to the centroid of the stars under
consideration $X_\odot,Y_\odot,X_\odot$; Figs. 5b, 6b, and 7b show
the parameters $M_{pq}$.

Analysis of these data leads us to conclude that: (a) when stars
with large distances are used, a significant bias is observed only
for the velocity of the Sun $Y_\odot$, while all three velocity
components of the Sun are subjected to a similar bias for a large
range of specified errors (Fig. 7a); (b) the parameters $M_{pq}$
being determined are smoothed out as the level of model errors
increases, which is particularly clearly seen from the middle part
of Fig. 7b; (c) it is necessary to reject very close ($d<200-250$
pc) stars for a reliable determination of all model parameters,
except the velocity components of the Sun, which, on the contrary,
are determined more reliably from the samples of nearby stars; (d)
three parameters are least subjected to any biases:
$M_{\scriptscriptstyle12}^{\scriptscriptstyle+}$,
 $M_{\scriptscriptstyle21}^{\scriptscriptstyle-}$ (the Oort constants), and
($M_{\scriptscriptstyle11}^{\scriptscriptstyle+}-M_{\scriptscriptstyle22}^{\scriptscriptstyle+}$).

We calculated the correlation coefficients (coefficient $k$)
between the parameters being determined, which arise because the
system of equations (7)--(8) is ill-conditioned:
 (i) $k=0.44$ for the pair $(M_{\scriptscriptstyle23}^{\scriptscriptstyle+})-(M_{\scriptscriptstyle32}^{\scriptscriptstyle-})$,
 (ii) $k=0.41$ for the pair $(M_{\scriptscriptstyle33}^{\scriptscriptstyle+}-M_{\scriptscriptstyle22}^{\scriptscriptstyle+})-
  (M_{\scriptscriptstyle11}^{\scriptscriptstyle+}-M_{\scriptscriptstyle22}^{\scriptscriptstyle+})$,
 (iii) for the sample of most distant stars
(Figs. 5--7), $k = 0.37$ for the pair
$(M_{\scriptscriptstyle12}^{\scriptscriptstyle+})-
  (M_{\scriptscriptstyle21}^{\scriptscriptstyle-})$, while for closer stars,
the correlation coefficient $k$ is nearly zero for this pair. The
correlations between the velocity $Y_\odot$, which is subjected to
the largest bias, and all the remaining model parameters are
nearly zero.

The slight increase in velocity $Y_\odot$ with distance that we
see in Fig. 5a can be explained by an asymmetric drift (Bobylev
and Bajkova 2007). Indeed, the mean value of $\Delta S$ changes
gradually from $0.1$ (${\overline d} = 150$ pc) to $-0.1$
(${\overline d} = 900$ pc).

\section{DISCUSSION}

Galactic Rotation Parameters As can be seen from the first column
of Table 2, the Oort constants are:
 $A = 15.9\pm0.2$ km s$^{-1}$ kpc$^{-1}$ and
 $B = -12.0\pm0.2$ km s$^{-1}$ kpc$^{-1}$.
Based on these parameters, we estimate the circular velocity of
rotation of the solar neighborhood to be
$V_0=|R_0\Omega_Z|=222\pm2$ km s$^{-1}$ for $R_0 = 8.0$ kpc and
the period of its revolution around the Galactic center to be 221
Myr.

The Oort constants found are in good agreement with their
determinations in other works. For example, using data on young
objects of the Galactic disk, Bobylev et al. (2008) found
 $A = 15.5\pm0.3$ km s$^{-1}$ kpc$^{-1}$ and
 $B =-12.2\pm0.7$ km s$^{-1}$ kpc$^{-1}$.

\subsection{Parameters of the Velocity Ellipsoid}

We clearly see from Table 2 that the kinematic parameters differ
for different groups of giants located near the Galactic plane at
$|Z| < 200$ pc (column 4) and at large elevations above the
Galactic plane $|Z|\geq 200$ pc (column 5). The vertex deviation
reaches $l_{xy} = 9.1\pm0^\circ.5$ for the first group and is a
factor of 2 smaller, $4.9\pm0^\circ.6,$ for the second group. Note
that the difference in kinematic parameters is enhanced with
increasing elevation above the Galactic plane. On the whole, our
results are in agreement with those of Rybka (2008).

From Table 3, we see an increase in velocity dispersions and a
decrease in vertex deviation with increasing ``age'' of the sample
of stars. Column no. 1 gives the parameters obtained for stars
with available radial velocities from all three space velocity
components. Using the radial velocities affects the determination
of the parameters of the velocity ellipsoid, with the most
prominent parameter being $l_1 = 11^\circ.7.$ For other samples,
there is also a difference in vertex deviation. Thus, for example,
$l_1 = 5^\circ.1$ for the sample of oldest stars is considerably
smaller than $l_1 = 9^\circ.4$ for the youngest group.

On the other hand, since RGC stars with various ages are well
mixed in space, using various constraints on the spatial
coordinates of stars does not lead to any noticeable change in the
parameters of the velocity ellipsoid (Table 4).

In contrast to the approach used by Rybka (2008), we traced the
changes in the parameters of the velocity ellipsoid with $|Z|$
(Fig. 2). As expected, our results show that younger stars with a
lower velocity dispersion are concentrated to the Galactic plane
to a greater extent.

\subsection{The Contraction Effect}

According to Table 2, the Oort constant K reaches its maximum
value for the sample of distant (500--1000 pc, $d = 0.65$ kpc,
column no. 4) RGC stars located near the Galactic plane ($|Z| <
200$ pc), $-5.3\pm1.3$ km s$^{-1}$ kpc$^{-1}$, while the
contraction velocity is $Kd = -3.5\pm0.9$ km s$^{-1}$; this value
is indicated in Fig. 4 by the filled circle. The open circles in
Fig. 4 indicate five values of $Kd$ from the cited data for which
estimates of the average distance are available. Below, we list
them, as denoted by the numbers in Fig. 4, in order of increasing
d:

(1) the contraction velocity determined from the space velocities
of A0--A5 giants (Bobylev et al. 2006b) for which
 $K = -13.1\pm2.0$ km s$^{-1}$ kpc$^{-1}$ was obtained, with $d=0.19$ kpc;

(2) the estimate from Rybka (2004b) obtained using a sample of
$\approx$11000 Tycho-2 G5--K0 giants with $|b|<50^\circ, d>250$
pc, $d=0.364$ kpc;

(3) the estimate from Rybka (2004a) obtained using a sample of ?
9000 main-sequence Tycho-2 B0--B9 stars, $d=0.53$ kpc;

(4) the estimate from Rybka (2008) obtained using $\approx$30000
Tycho-2 RGC stars at $|b|<30^\circ$ and $d > 100$ pc for which we
found the average photometric distance based on the catalog by
Gontcharov (2008), $d=0.54$ kpc;

(5) the result obtained using open star clusters with an average
age of $\approx$100 Myr (Bobylev et al. 2007) at $d=1.05$ kpc.

Note also the result of analyzing OB stars located in the range of
distances 600--2000 pc from Torra et al. (2000) for which
 $K = -2.9\pm0.6$ km s$^{-1}$ kpc$^{-1}$ was found. Taking ${\overline
d}\approx0.8$ kpc as a realistic average distance for these stars
(which is not given in Torra et al. (2000)), we will obtain
 $Kd =-2.3$ km s$^{-1}$, a value consistent with the data in Fig. 4. A
comprehensive overview of the results of determining the negative
K effect by other authors can be found in Fern\'andez et al.
(2001).

As can be seen from Fig. 4, on the whole, there is good agreement
between the various contraction velocity estimates. The main
purpose of our comparison is to show that the results of different
authors obtained by different methods agree in sign, i.e., the
contraction effect actually takes place, and, given the errors,
they also agree satisfactorily in magnitude. However, the
difference for the A0--A5 giants of the OSACA database (Bobylev et
al. 2006b) is most pronounced. This difference can be explained by
the fact that the sample of A0--A5 stars was specially formed by a
kinematic indicator傭y the maximum K effect. At the same time,
this result is in good agreement with the data of Table 1,
according to which we have $Kd = -1.8\pm0.5$ km s$^{-1}$.

What is the maximum contraction velocity? In the opinion of Rybka
(2004a), $Kd = -8.2\pm0.9$ km s$^{-1}$ at $d = 1$ kpc. However,
this estimate was obtained by extrapolating the data on
main-sequence dwarfs ($d<0.6$ kpc). Analysis of the kinematics of
open star clusters (Bobylev et al. 2007) showed the magnitude of
the velocity Kd to decrease at $d>1$ kpc.

Note also the ambiguity of treating $K=0.5(M_{11}+M_{22})$ as the
effect of contraction in the plane. Indeed, as we see from Table
1, $M_{22}=\partial u_2/\partial r_2$ does not differ
significantly from zero, with large values of $K$ and $l_{xy}$
being reached solely through the gradient $M_{11}=\partial
u_1/\partial r_1$. Thus, we can only assert that noticeable
deviations from circular motions are observed. The nature of these
deviations requires a further study.

%\end{document}
%%%%%%%%%%%%%%%%%%

\subsection{The Problem of the Galactic Disk Warp}

{\bf Rotation around the X axis.} As can be seen from Table 2 (the
third column),
 $M_{\scriptscriptstyle32}^{\scriptscriptstyle-}=-1.4\pm0.2$ km s$^{-1}$ kpc$^{-1}$
 and
 $M_{\scriptscriptstyle23}^{\scriptscriptstyle+}=+1.1\pm0.3$ km s$^{-1}$ kpc$^{-1}$. The
negative sign of
 $M_{\scriptscriptstyle32}^{\scriptscriptstyle-}$ means that the rotation is from the $Z$ axis
to the $Y$ axis, i.e., the stars approach to the Galactic plane.
They approach from the North Pole in the first and second
quadrants and from the South Pole in the third and fourth
quadrants. By analogy with the rotation around the
$Z(\Omega_Z=B-A)$ axis, the angular velocity of rotation around
the $X$ axis can be determined as
 $\Omega_X=M_{\scriptscriptstyle32}^{\scriptscriptstyle-}-
           M_{\scriptscriptstyle23}^{\scriptscriptstyle+}=-2.5\pm0.3$ km s$^{-1}$ kpc$^{-1}$.

This is inconsistent with the analysis of the motions of
$\approx$4000 giants of various spectral types O--M performed by
Miyamoto et al. (1993), who found a positive direction of rotation
around the $X$ axis (directed toward the Galactic center). Note
that, in this case, the stellar proper motions being analyzed were
determined in the FK5 system, which is noticeably distorted by the
uncertainty in the precession constant. On the other hand, having
analyzed the proper motions of O--B5 stars with the Hipparcos
(19977) system, Miyamoto and Zhu (1998) also reached the
conclusion about a positive rotation around the $X$ axis. In our
notation, Miyamoto and Zhu (1998) obtained
 $\Omega_X=2M_{\scriptscriptstyle32}^{\scriptscriptstyle-}\equiv
  2M_{\scriptscriptstyle23}^{\scriptscriptstyle+}=+3.8\pm1.1$ km s$^{-1}$ kpc$^{-1}$.
 However, it should be noted that
Miyamoto et al. (1993) and Miyamoto and Zhu (1998) used a model
where the variables
 $M_{\scriptscriptstyle32}^{\scriptscriptstyle-}$ and
 $M_{\scriptscriptstyle23}^{\scriptscriptstyle+}$ could not be separated and,
therefore, they were assumed to be equivalent. In our approach,
which consists in solving Eqs. (7)--(8), we can separate these
variables. This is possible, because the distribution of RGC stars
over the celestial sphere is fairly uniform.

The parameters found with a constraint on $|Z|$ and given in
column no. 4 of Table 2 are also of great interest. As we see, for
the sample of stars close to the Galactic plane (this case
corresponds largely to the distribution of O--B5 stars), the
solutions for M with index 3 differ markedly from those obtained
for the samples filling uniformly the celestial sphere (Table 2,
columns nos. 1, 2, 3, 5).

As can be seen from Table 2 and Fig. 5, the values of the
parameters $M^-_{32}$ and $M^+_{23}$ do not depend strongly on the
distance. However, the motions of nearby stars can be distorted by
local peculiarities (in particular, by the kinematic peculiarities
of the Gould Belt with a characteristic radius of about 500 pc).
Therefore, the values of these parameters determined from distant
stars seem more reliable. Thus, we interpret $\Omega_X$ found from
relatively distant RGC stars as the rotation related to the warp
of the Galactic stellar-gaseous disk.

Note that owing to the use of currently available data that
realize the inertial frame of reference more accurately than FK5
and the possibility of separating the components $M^-_{32}$ and
$M^+_{23}$ when solving Eqs. (7)--(8), we have every reason to
believe that our results are more reliable than those of Miyamoto
et al. (1993) and Miyamoto and Zhu (1998).

{\bf Rotation around the Y axis.}
 There is a large uncertainty
with regard to the magnitude and direction of the rotation around
the Y axis related to the disk warp. Thus, for example, having
analyzed the motions of $\approx$4000 giants of various spectral
types O--M, Miyamoto et al. (1993) found a positive direction of
rotation around the $Y$ axis:
 $\Omega_Y=2M_{\scriptscriptstyle13}^{\scriptscriptstyle-}\equiv
 -2M_{\scriptscriptstyle13}^{\scriptscriptstyle+}=+5.6\pm1.0$ km s$^{-1}$ kpc$^{-1}$.
 Based on an analysis of the Hipparcos data, Drimmel et
al. (2000) obtained a completely different estimate in the form of
precession of OB stars in the $ZX$ plane with an angular velocity
of $-25$ km s$^{-1}$ kpc$^{-1}$ (i.e., rotation around the $Y$
axis). However, Bobylev (2004) showed that this value could be
explained by inaccurate referencing of the ICRS/Hipparcos
realization to the system of extragalactic sources.

Based on the data of column no. 2 in Table 2, we have
 $\Omega_Y=M_{\scriptscriptstyle13}^{\scriptscriptstyle-}
 -M_{\scriptscriptstyle13}^{\scriptscriptstyle+}=-1.5\pm0.3$ km s$^{-1}$ kpc$^{-1}$.
 As we see, the value of this
component is determined almost entirely by $M^-_{13}$. In this
respect,
 $M_{\scriptscriptstyle13}^{\scriptscriptstyle-}/4.74=-0.41\pm0.05$ mas yr$^{-1}$,
 which is in
excellent agreement with the residual rotation of the ICRS
relative to the inertial frame of reference,
 $M_{\scriptscriptstyle13}^{\scriptscriptstyle-}=-0.37\pm0.04$ mas yr$^{-1}$,
 estimated by Bobylev and Khovrichev (2006) based on
the Tycho-2 and UCAC2 (Zacharias 2004) Catalogues, is of great
interest. Therefore, we cannot say what the contribution related
to the disk warp is in this case, since these two effects cannot
be separated. Obviously, if we apply the correction
 $M_{\scriptscriptstyle13}^{\scriptscriptstyle-}\approx-0.4$ mas yr$^{-1}$
 (reducing the Hipparcos system to the inertial frame of
reference), then the ``remainder'' of the rotation around $Y$ axis
will be close to zero.

On the whole, we may conclude that the kinematic effect related to
the disk warp manifests itself mainly as the rotation around the
$X$ axis (see the previous section). The small magnitude of the
rotation velocity around the $Y$ axis is in agreement with the
fact that the line of nodes of the HI layer is known to be close
to the direction of the Galactic center/anticenter.

\section*{CONCLUSIONS}

We analyzed the proper motions of 95633 RGC stars from the Tycho-2
Catalogue using the linear Ogorodnikov-Milne model. The
photometric distances to these stars were estimated by Gontcharov
(2008) based on 2MASS photometric data.

Based on a subsample of 3632 RGC stars for which not only the
proper motions but also the radial velocities, photometric
distances, and Hipparcos trigonometric parallaxes are available,
we showed that, apart from the star centroid velocity components
relative to the Sun $(X_\odot,Y_\odot,Z_\odot),$ only those
parameters of the Ogorodnikov-Milne model that describe the
stellar motions in the $XY$ plane differ significantly from zero.
This allowed us to study the plane $K$ effect based the tangential
velocities of a considerably larger sample of RGC stars.

To ascertain the dependence of the kinematic parameters being
determined on the distances, including the measurement errors in
the distances, we performed statistical Monte Carlo simulations.
Our simulations showed that, first, a significant bias is observed
only for the velocity $Y_\odot$ at large distances and, second,
very close ($d<200-250$ pc) stars should be excluded to reliably
determine all parameters of the Ogorodnikov-Milne model, except
the velocity components of the Sun, which, on the contrary, are
determined more reliably from samples of nearby stars.

Analysis of the entire sample of RGC stars showed that the Oort
constants that describe the rotation around the Z axis are
 $A = 15.9\pm0.2$ km s$^{-1}$ kpc$^{-1}$ and $B = -12.0\pm0.2$ km s$^{-1}$ kpc$^{-1}$.
 This gives an estimate of the rotation velocity of the solar
neighborhood, $V_0 = 222\pm2$ km s$^{-1}$, for $R_0 = 8.0$ kpc and
the period of its revolution around the Galactic center, 221 Myr.

Judging by the velocity dispersions found, the RGC sample includes
stars of various ages.We made an attempt to roughly separate the
RGC stars by ages based on the parameter $\Delta S$. It showed
that only the fraction of relatively young RGC stars, which
accounts for about 20\% of the entire sample, is identified most
reliably.

The most interesting results were obtained using stars located in
the range of distances 500--1000 pc. The RGC stars located near
the Galactic plane were shown to have a noticeable contraction
effect (observed in the XY plane) that depends on the heliocentric
distances of the stars. Thus, for example, for the sample of RGC
stars located near the Galactic plane ($|Z|<200$ pc), the
contraction velocity is $Kd = -3.5\pm0.9$ km s$^{-1}$ and the
vertex deviation is $l_{xy} = 9.1\pm0^\circ.5$. For stars located
well above the Galactic plane ($|Z|\geq200$ pc), these effects are
less pronounced, being $Kd = -1.7\pm0.5$ km s$^{-1}$ and $l_{xy}=
4.9\pm0^\circ.6$.

Based on RGC stars with distances in the range 500--1000 pc, we
found the angular velocity of rotation around the Galactic $X$
axis to be $-2.5\pm0.3$ km s$^{-1}$ kpc$^{-1}$, which we interpret
as the rotation related to the warp of the Galactic
stellar-gaseous disk. The direction of this rotation shows that
the stars approach the Galactic plane.

\section*{ACKNOWLEDGMENTS}

We are grateful to the referees for helpful remarks that
contributed to a significant improvement of the paper. This work
was supported by the Russian Foundation for Basic Research
(project no. 08--02--00400) and in part by the ``Origin and
Evolution of Stars and Galaxies'' Program of the Presidium of the
Russian Academy of Sciences and the Program for State Support of
Leading Scientific Schools of Russia (NSh-6110.2008.2
``Multiwavelength Astrophysical Research'').

 \bigskip
{\bf REFERENCES}
 \bigskip

 1. W.B. Burton, Galactic and Extragalactic Radio
Astronomy, Ed. by G. Verschuur and K. Kellerman (Springer, New
York, 1988).

2. V.V. Bobylev, Pis知a Astron. Zh. 30, 185 (2004) [Astron. Lett.
30, 159 (2004)].

3. V.V. Bobylev and A. T. Bajkova, Astron. Zh. 84, 418 (2007)
[Astron. Rep. 51, 372 (2007)].

4. V.V. Bobylev, A.T. Bajkova, and G.A. Gontcharov, Astron.
Astrophys. Trans. 25, 143 (2006a).

5. V.V. Bobylev, A.T. Bajkova, S.V. Lebedeva, Pis知a Astron. Zh.
33, 809 (2007) [Astron. Lett. 33, 720 (2007)].

6. V.V. Bobylev, A.T. Bajkova, and A.S. Stepanishchev, Pis知a
Astron. Zh. 34, 570 (2008) [Astron. Lett. 34, 515 (2008)].

7. V.V. Bobylev, G.A. Gontcharov, and A.T. Bajkova, Astron. Zh.
83, 821 (2006b) [Astron. Rep. 50, 733 (2006)].

8. V.V. Bobylev and M. Yu. Khovrichev, Pis知a Astron. Zh. 32, 676
(2006) [Astron. Lett. 32, 608 (2006)].

9. S.V.M. Clube, Mon. Not. R. Astron. Soc. 159, 289 (1972).

10. S.V.M. Clube, Mon. Not. R. Astron. Soc. 161, 445 (1973).

11. W. Dehnen, Astron. J. 115, 2384 (1998).

12. R. Drimmel, R.L. Smart, and M.G. Lattanzi, Astron. Astrophys.
354, 67 (2000).

13. R. Drimmel and D.N. Spergel, Astroph. J. 556, 181 (2001).

14. D. Fern\'andez, F. Figueras, and J. Torra, Astron. Astrophys.
372, 833 (2001).

15. G.A. Gontcharov, Pis知a Astron. Zh. 32, 844 (2006) [Astron.
Lett. 32, 759 (2006)].

16. G.A. Gontcharov, Pis知a Astron. Zh. 34, 868 (2008) [Astron.
Lett. 34, 785 (2008)].

17. The HIPPARCOS and Tycho Catalogues, ESA SP- 1200 (1997).

18. E. Hog, C. Fabricius, V.V. Makarov, et al., Astron. Astrophys.
355, L27 (2000).

19. P.M.W. Kalberla and L. Dedes, Astron. Astrophys. 487, 951
(2008).

20. M.R. Metzger, J.A.R. Caldwell and P.L. Schechter, Astron. J.
115, 635 (1998).

21. M. Miyamoto, M. Soma, and M. Yoshizawa, Astron. J. 105, 2138
(1993).

22. M. Miyamoto and Z. Zhu, Astron. J. 115, 1483 (1998).

23. K.F. Ogorodnikov, Dynamics of Stellar Systems (Fizmatgiz,
Moscow, 1965) [in Russian].

24. R.P. Olling and W. Dehnen, Astroph. J. 599, 275 (2003).

25. P.P. Parenago, Tr. GAISh 20, 26 (1951).

26. P.P. Parenago, A Course on Stellar Astronomy (Gosizdat,
Moscow, 1954) [in Russian].

27. F. Pont, M. Mayor, and G. Burki, Astron. Astrophys. 285, 415
(1994).

28. K. Rohlfs, Lectures in Density Waves (Springer, Berlin, 1977).

29. S.P. Rybka, Kinem. Fiz. Neb. Tel 20, 133 (2004a).

30. S.P. Rybka, Kinem. Fiz. Neb. Tel 20, 437 (2004b).

31. S.P. Rybka, Kinem. Fiz. Neb. Tel 22, 225 (2006).

32. S.P. Rybka, Kinem. Fiz. Neb. Tel 23, 102 (2007) [Kin. Phys.
Cel. Bodies 23, 70 (2007)].

33. S.P. Rybka, Kinem. Fiz. Neb. Tel 24, 137 (2008) [Kin. Phys.
Cel. Bodies 23, 99 (2008)].

34. M.F. Skrutskie, R.M. Cutri, R. Stiening, et al., Astron. J.
131, 1163 (2006).

35. M. Steinmetz, T. Zwitter, A. Seibert, et al., Astron. J. 132,
1645 (2006).

36. J. Torra, D. Fern\'andez, and F. Figueras, Astron. Astrophys.
359, 82 (2000).

37. R.J. Trumpler and H.F. Weaver, Statistical Astronomy (Univ. of
Calif., Berkely, 1953).

38. I. Yusifov, astro-ph: 0405517 (2004).

39. N. Zacharias, S.E. Urban, M.I. Zacharias, et al., Astron. J.
127, 3043 (2004).

\newpage
%%%%%%%%%%%%%%%%%%%%%%%%%%%%%%%%%%%%%%%%%%%%%%
{
\begin{table}[p]                                                %% t-1
\caption[]{\small\baselineskip=1.0ex\protect Kinematic parameters
obtained by solving the system of equations (3)--(5)
 }
\begin{center}
 \small
\begin{tabular}{|r|r|r|r|r}\hline
       $N_\star$    &    3632       &   3632        \\\hline
                    &    1~~~~      &   2~~~~       \\\hline
    $X_\odot$,~km s$^{-1}$ & $  9.9\pm0.4$ & $  9.5\pm0.4$ \\
    $Y_\odot$,~km s$^{-1}$ & $ 16.8\pm0.4$ & $ 16.3\pm0.4$ \\
    $Z_\odot$,~km s$^{-1}$ & $  7.3\pm0.4$ & $  6.9\pm0.4$ \\
 $M_{11}$,~km s$^{-1}$ kpc$^{-1}$ & $-15.5\pm3.3$ & $-14.3\pm3.7$ \\
 $M_{12}$,~km s$^{-1}$ kpc$^{-1}$ & $ 28.7\pm3.2$ & $ 29.3\pm3.5$ \\
 $M_{13}$,~km s$^{-1}$ kpc$^{-1}$ & $ -2.8\pm2.6$ & $  0.6\pm2.9$ \\
 $M_{21}$,~km s$^{-1}$ kpc$^{-1}$ & $  8.6\pm3.3$ & $  5.9\pm3.7$ \\
 $M_{22}$,~km s$^{-1}$ kpc$^{-1}$ & $ -1.3\pm3.2$ & $ -2.2\pm3.5$ \\
 $M_{23}$,~km s$^{-1}$ kpc$^{-1}$ & $ -5.8\pm2.6$ & $ -3.8\pm2.9$ \\
 $M_{31}$,~km s$^{-1}$ kpc$^{-1}$ & $ -6.7\pm3.3$ & $ -4.3\pm3.7$ \\
 $M_{32}$,~km s$^{-1}$ kpc$^{-1}$ & $ -5.3\pm3.2$ & $ -4.1\pm3.5$ \\
 $M_{33}$,~km s$^{-1}$ kpc$^{-1}$ & $ -1.1\pm2.6$ & $ -4.3\pm2.9$ \\
      $A$,~km s$^{-1}$ kpc$^{-1}$ & $ 18.6\pm2.3$ & $ 17.6\pm2.5$ \\
      $B$,~km s$^{-1}$ kpc$^{-1}$ & $-10.0\pm2.3$ & $-11.7\pm2.5$ \\
      $C$,~km s$^{-1}$ kpc$^{-1}$ & $ -7.1\pm2.3$ & $ -6.1\pm2.5$ \\
      $K$,~km s$^{-1}$ kpc$^{-1}$ & $ -8.4\pm2.3$ & $ -8.2\pm2.5$ \\
 ${\overline d}$,~kpc &      $0.22$ &        $0.22$ \\
 \hline
\end{tabular}
\end{center}
 \small\baselineskip=1.0ex\protect
 Note. The parameters presented in columns nos. 1 and 2 were
obtained using trigonometric parallaxes and photometric distances,
respectively; $N_\star$ is the number of sample stars.

\vskip50mm
 \end{table}
}

%%%%%%%%%%%%%%%%%%%%%%%%%%%%%%%%%%%%%%%%%%%%%%
%%%%%%%%%%%%%%%%%%%%%%%%%%%%%%%%%%%%%%%%%%%%%%
\newpage
%%%%%%%%%%%%%%%%%%%%%%%%%%%%%%%%%%%%%%%%%%%%%%
{
\begin{table}[t]                                                %% t-2
\caption[]{\small\baselineskip=1.0ex\protect
 Kinematic parameters obtained by solving the system of
equations (7)--(8)
 }
\begin{center}
 \tiny
\begin{tabular}{|r|r|r|r|r|r|}\hline
                          &  $d<1000$~pc   & $500-1000$~pc  &  $700-1000$~pc & $500-1000$~pc  & $500-1000$~pc   \\
                          &                &                &                &  $|Z|<200$~pc  &  $|Z|\geq200$~pc \\
             $N_\star$    &     95633~~    &   50595~~      &      19200~~   &  22684~~       &   27911~~     \\\hline
                          &       1~~~~    &     2~~~~      &        3~~~~   &    4~~~~       &     5~~~~     \\\hline
          $X_\odot$,~km s$^{-1}$ & $ 7.84\pm0.09$ & $ 8.55\pm0.13$ & $ 8.67\pm0.22$ & $ 8.49\pm0.18$ & $ 8.52\pm0.18$\\
          $Y_\odot$,~km s$^{-1}$ & $15.94\pm0.10$ & $17.74\pm0.14$ & $19.19\pm0.26$ & $16.14\pm0.23$ & $18.48\pm0.19$\\
          $Z_\odot$,~km s$^{-1}$ & $ 6.60\pm0.08$ & $ 7.17\pm0.12$ & $ 7.59\pm0.21$ & $ 6.86\pm0.15$ & $ 7.59\pm0.19$\\

$M_{21}^{+}(A)$,~km/s/kpc & $15.87\pm0.20$ & $15.86\pm0.24$ & $15.61\pm0.37$ & $16.25\pm0.35$ & $15.33\pm0.35$\\
   $M_{32}^{-}$,~km/s/kpc & $-1.40\pm0.17$ & $-1.41\pm0.21$ & $-1.80\pm0.31$ & $ 1.83\pm0.95$ & $-1.70\pm0.26$\\
   $M_{13}^{-}$,~km/s/kpc & $-2.01\pm0.18$ & $-1.93\pm0.22$ & $-1.97\pm0.32$ & $-5.28\pm0.77$ & $-1.70\pm0.27$\\
$M_{21}^{-}(B)$,~km/s/kpc &$-11.99\pm0.15$ &$-11.99\pm0.18$ &$-12.18\pm0.26$ &$-11.82\pm0.24$ &$-12.26\pm0.27$\\
$M_{11-22}^{+}$,~km/s/kpc & $-7.86\pm0.38$ & $-7.43\pm0.45$ & $-6.47\pm0.68$ &$-10.73\pm0.62$ & $-5.29\pm0.66$\\
   $M_{13}^{+}$,~km/s/kpc & $-0.45\pm0.23$ & $-0.47\pm0.27$ & $ 0.15\pm0.40$ & $-3.90\pm0.82$ & $-0.31\pm0.38$\\
   $M_{23}^{+}$,~km/s/kpc & $ 0.98\pm0.22$ & $ 1.13\pm0.26$ & $ 1.49\pm0.38$ & $-1.79\pm1.01$ & $ 0.88\pm0.36$\\
$M_{33-22}^{+}$,~km/s/kpc & $-1.36\pm0.43$ & $-0.99\pm0.51$ & $-0.47\pm0.75$ & $-0.02\pm1.27$ & $-0.27\pm0.61$\\

     ${\overline d}$,~kpc & $ 0.52\pm0.21$ & $ 0.68\pm0.14$ & $ 0.81\pm0.11$ & $ 0.65\pm0.12$ & $0.70\pm0.15$\\
            $C$,~km s$^{-1}$ kpc$^{-1}$ & $ -3.9\pm0.2$  & $ -3.7\pm0.2$  & $ -3.2\pm0.3$  & $ -5.4\pm0.3$  & $ -2.7\pm0.3$\\
            $K$,~km s$^{-1}$ kpc$^{-1}$ & $ -2.6\pm0.5$  & $ -2.7\pm0.6$  & $ -2.8\pm0.8$  & $ -5.3\pm1.3$  & $ -2.4\pm0.7$\\
      $K\cdot d$,~km s$^{-1}$ & $ -1.4\pm0.3$  & $ -1.8\pm0.3$  & $ -2.3\pm0.7$  & $ -3.5\pm0.9$  & $ -1.7\pm0.5$\\
               $l_{xy}$, deg. & $  7.0\pm0.3$  & $  6.6\pm0.4$  & $  5.9\pm0.6$  & $  9.1\pm0.5$  & $  4.9\pm0.6$\\
 \hline
\end{tabular}

\end{center}
\vskip40mm
\end{table}
}

{ %%%%%%%%%%%%%%%%%%%%%%%%%%%%%%%%%%%%%%%%%%%%%%%%%%%              t-3
\begin{table}[p]
\caption[]{\small\baselineskip=1.0ex\protect
 Parameters of the residual velocity ellipsoid
 }
\begin{center}
 \small
\begin{tabular}{|r|r|r|r|r|r|}\hline
      $N_\star$ &      3632        &     39979     &   36529       &  20497        & $\varepsilon~$ \\\hline
                &       1~~        &        2~~    &     3~~       &    4~~        &       5~       \\\hline
     $\sigma_1$ &   $32.1\pm0.4$~  &    $29.1$     &    $35.6$     & $19.7$        &      $0.50$     \\
     $\sigma_2$ &   $21.2\pm0.4$~  &    $21.7$     &    $25.9$     & $13.1$        &      $0.40$     \\
     $\sigma_3$ &   $17.9\pm0.3$~  &    $16.8$     &    $20.6$     & $10.8$        &      $0.25$     \\
 $l_1$ & $ 11^\circ.7\pm4^\circ.0$ & $  6^\circ.3$ & $  5^\circ.1$ & $  9^\circ.4$ & $ 0^\circ.4$ \\
 $l_2$ & $101^\circ.7\pm1^\circ.4$ & $ 96^\circ.3$ & $ 91^\circ.8$ & $100^\circ.1$ & $ 0^\circ.4$ \\
 $l_3$ & $287^\circ.6\pm1^\circ.4$ & $260^\circ~~$ & $253^\circ~~$ & $265^\circ~~$ & $ 15^\circ~~$ \\
 $b_1$ & $ -4^\circ.1\pm0^\circ.8$ & $  1^\circ.0$ & $  0^\circ.8$ & $  3^\circ.0$ & $ 0^\circ.2$ \\
 $b_2$ & $  3^\circ.9\pm0^\circ.8$ & $  3^\circ.6$ & $  5^\circ.0$ & $ 11^\circ.9$ & $ 0^\circ.6$ \\
 $b_3$ & $ 86^\circ.1\pm1^\circ.8$ & $ 86^\circ.3$ & $ 87^\circ.9$ & $ 77^\circ.7$ & $ 0^\circ.5$ \\

 \hline
\end{tabular}
\end{center}
 \small\baselineskip=1.0ex\protect
 Note. Columns nos. 1 and 2--4 give the parameters determined from
the stellar space velocities and only from the stellar proper
motions, respectively; column no. 2 for $0<\Delta S,$ column no. 3
for $-0.38<\Delta S<0,$ column no. 4 for $-1<\Delta S<-0.38$;
column no. 5 gives a Monte Carlo estimate of the error
$\varepsilon$.

\vskip50mm
 \end{table}
}
%%%%%%%%%%%%%%%%%%%%%%%%%%%%%%%%%%%%%%%%%%%%%%
{
\begin{table}[p]                                                %% t-4
\caption[]{\small\baselineskip=1.0ex\protect
 Parameters of the residual velocity ellipsoid obtained
using the proper motions of RGC stars for the samples of Table 2
 }
\begin{center}
 \small
\begin{tabular}{|r|r|r|r|r|r|}\hline
            & $d<1000$~pc   & $500-1000$~pc & $700-1000$~pc & $500-1000$~pc & $500-1000$~pc   \\
            &               &              &              & $|Z|<200$~pc  & $|Z|\geq200$~pc \\
  $N_\star$ &     95633~~   &   50595~~    &     19200~~  & 22684~~       & 27911~~      \\\hline
            &       1~~~~   &     2~~~~    &       3~~~~  &   4~~~~       &   5~~~~      \\\hline
 $\sigma_1$ & $30.1$        &    $32.2$    &    $35.0$    &    $31.4$     & $33.0$       \\
 $\sigma_2$ & $21.7$        &    $23.2$    &    $25.3$    &    $20.3$     & $24.2$       \\
 $\sigma_3$ & $17.2$        &    $18.7$    &    $20.7$    &    $17.3$     & $20.4$       \\
 $l_1$      & $  6^\circ.0$ & $ 5^\circ.2$ & $ 5^\circ.5$ & $  6^\circ.6$ & $ 3^\circ.8$ \\
 $l_2$      & $ 96^\circ.0$ & $95^\circ.2$ & $95^\circ.5$ & $ 96^\circ.6$ & $93^\circ.8$ \\
 $l_3$      & $257^\circ.9$ & $91^\circ.1$ & $83^\circ.7$ & $242^\circ.7$ & $80^\circ.1$ \\
 $b_1$      & $  0^\circ.7$ & $-3^\circ.0$ & $-1^\circ.2$ & $  0^\circ.5$ & $-1^\circ.1$ \\
 $b_2$      & $  2^\circ.3$ & $-4^\circ.2$ & $-5^\circ.7$ & $  0^\circ.7$ & $-4^\circ.6$ \\
 $b_3$      & $ 87^\circ.6$ & $85^\circ.8$ & $84^\circ.2$ & $ 89^\circ.1$ & $85^\circ.3$ \\\hline
\end{tabular}
\end{center}
\end{table}
}
%%%%%%%%%%%%%%%%%%%%%%%%%%%%%%%%%%%%%%%%%%%%%%

\newpage
%%%%%%%%%%%%%%%%%%%%%%%%%%%%%%%%%%%%%%%% FIG.1:
\begin{figure}[p]
{
\begin{center}
 \includegraphics[width=120mm]{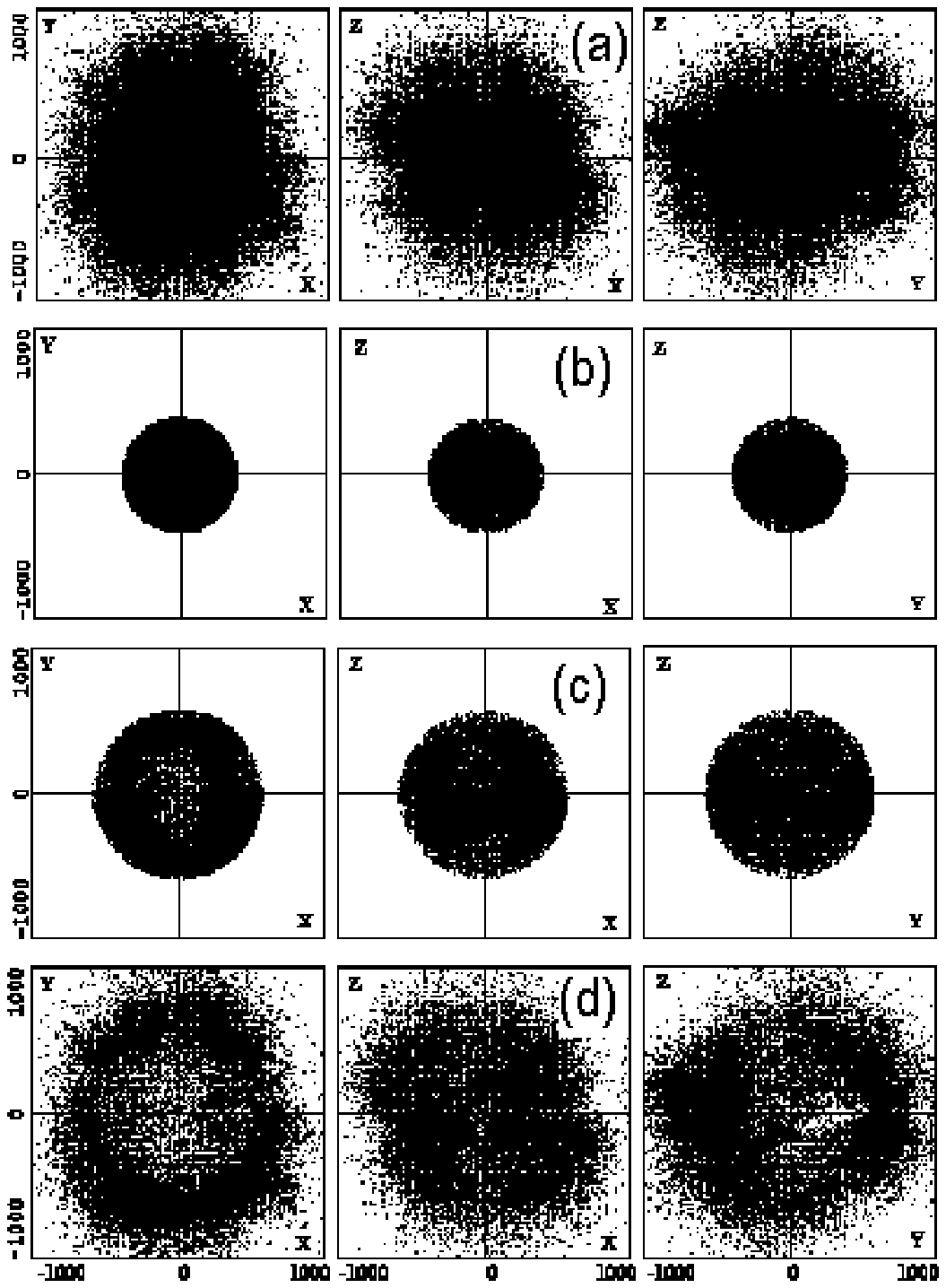}
\end{center}
} Fig.~1. Distributions of the samples of RGC stars in the $XY,$
$XZ,$ and $YZ$ planes (the left, central, and right columns,
respectively): the complete sample (a), the stars with distances
$d<400$ pc (b), 400 pc$<d<600$ pc (c), and 600 pc$<d<1600$ pc (d).
\end{figure}
%%%%%%%%%%%%%%%%%%%%%%%%%%%%%%%%%%%%%%%%%%%%%%%%%%%%%%%%%%%%%%%

\newpage
%%%%%%%%%%%%%%%%%%%%%%%%%%%%%%%%%%%%%%%% FIG.2:
\begin{figure}[t]
{
\begin{center}
 \includegraphics[width=80mm]{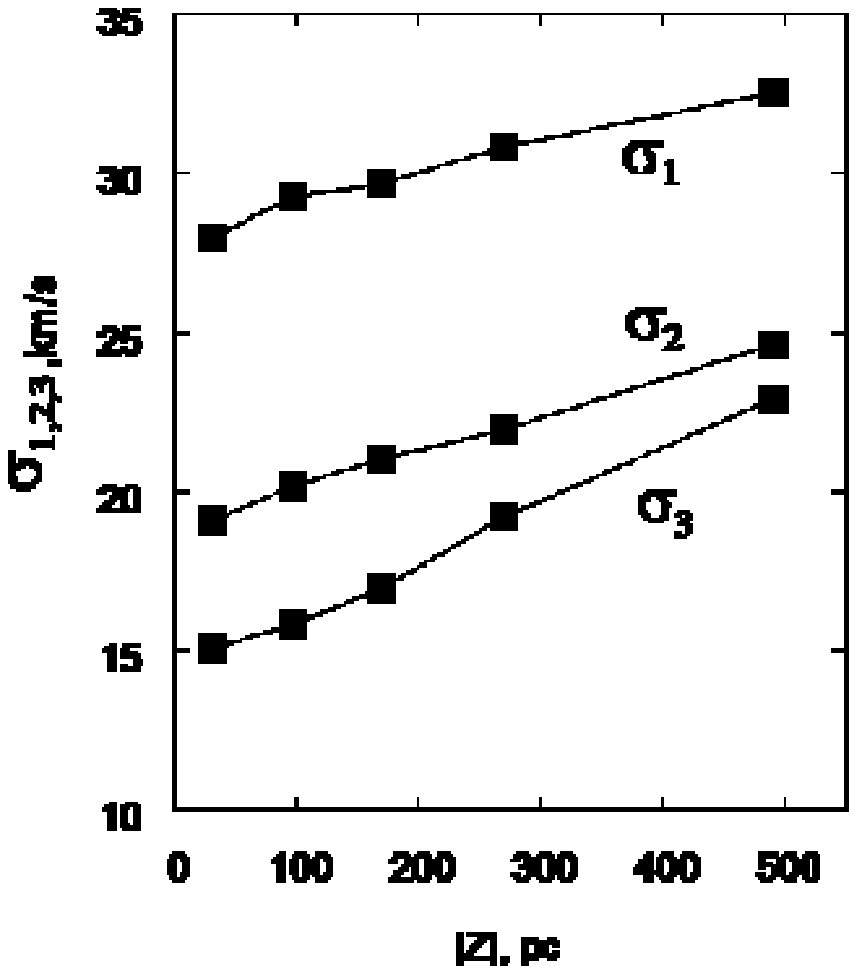}
\end{center}
} Fig.~2. Principal semiaxes of the residual velocity ellipsoid
$\sigma_1,\sigma_2,\sigma_3$ versus $|Z|$.
\end{figure}
%%%%%%%%%%%%%%%%%%%%%%%%%%%%%%%%%%%%%%%%%%%%%%%%%%%%%%%%%%%%%%%

\newpage
%%%%%%%%%%%%%%%%%%%%%%%%%%%%%%%%%%%%%%%% FIG.3:
\begin{figure}[t]
{
\begin{center}
 \includegraphics[width=100mm]{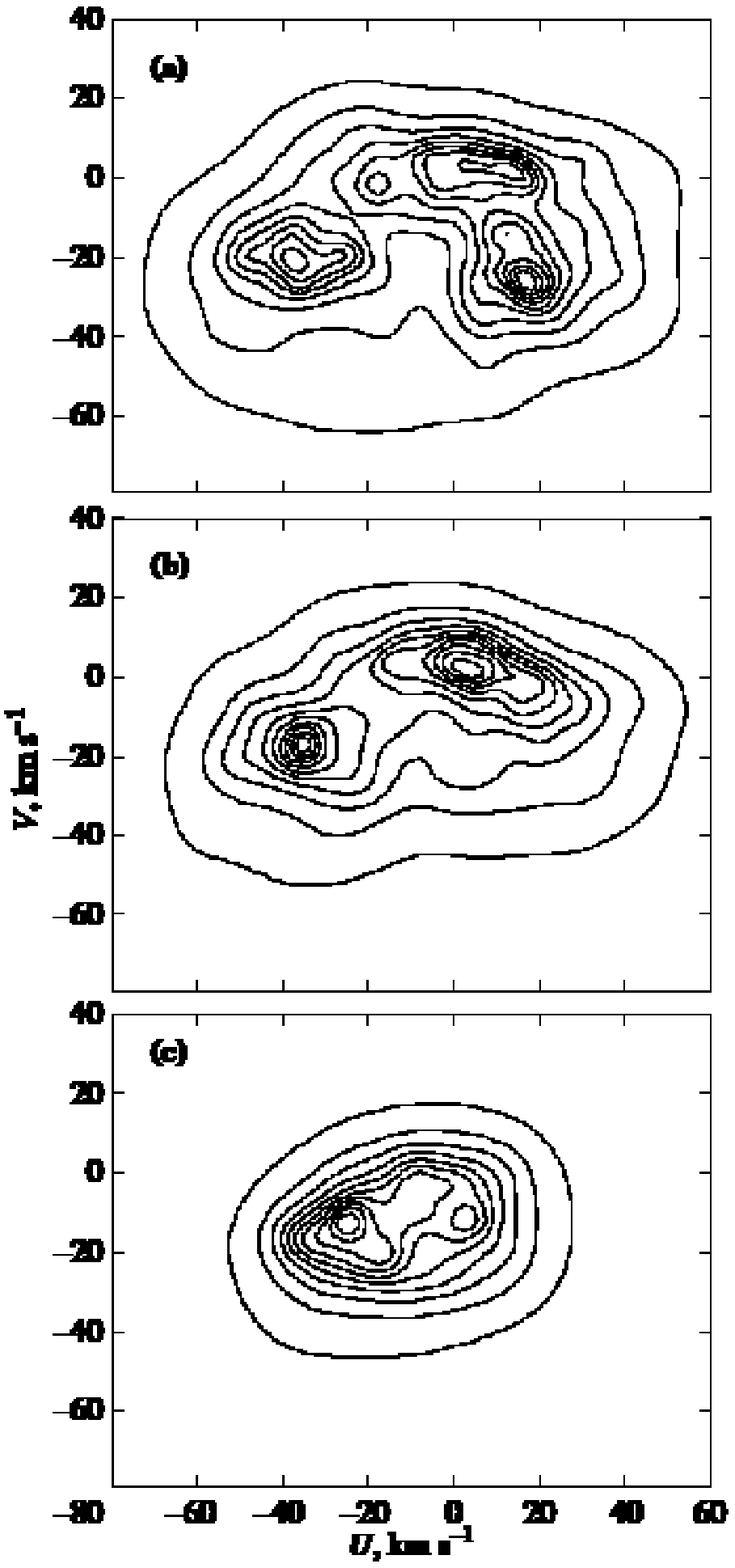}
\end{center}
} Fig.~3. Smoothed distributions of UV velocities: (a) sample (1)
for $0<\Delta S$; (b) sample (2) for $-0.38< \Delta S<0$; (c)
sample (3) for $-1<\Delta S< -0.38$ (see the text).
\end{figure}
%%%%%%%%%%%%%%%%%%%%%%%%%%%%%%%%%%%%%%%%%%%%%%%%%%%%%%%%%%%%%%%

\newpage
%%%%%%%%%%%%%%%%%%%%%%%%%%%%%%%%%%%%%%%% FIG.4:
\begin{figure}[t]
{
\begin{center}
 \includegraphics[width=120mm]{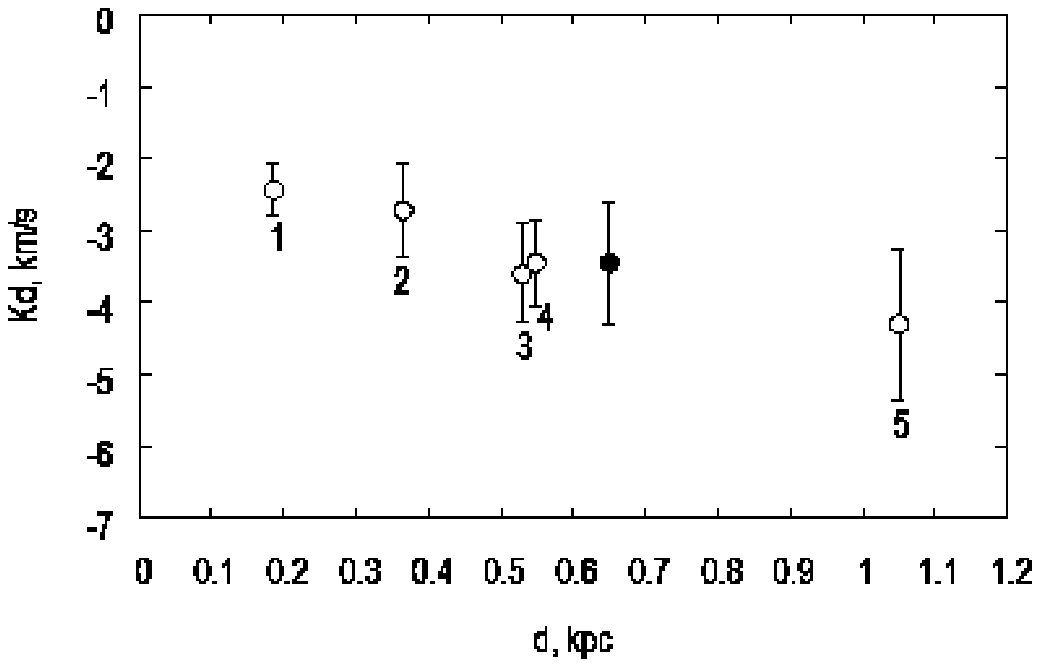}
\end{center}
} Fig.~4. Velocity $Kd$ versus heliocentric distance $d$; the open
circles mark the cited data (see the text) and the filled circles
indicate the value of $Kd$ from Table 2.
\end{figure}
%%%%%%%%%%%%%%%%%%%%%%%%%%%%%%%%%%%%%%%%%%%%%%%%%%%%%%%%%%%%%%%

\newpage
%%%%%%%%%%%%%%%%%%%%%%%%%%%%%%%%%%%%%%%% FIG.5:
\begin{figure}[t]
{
\begin{center}
 \includegraphics[width=100mm]{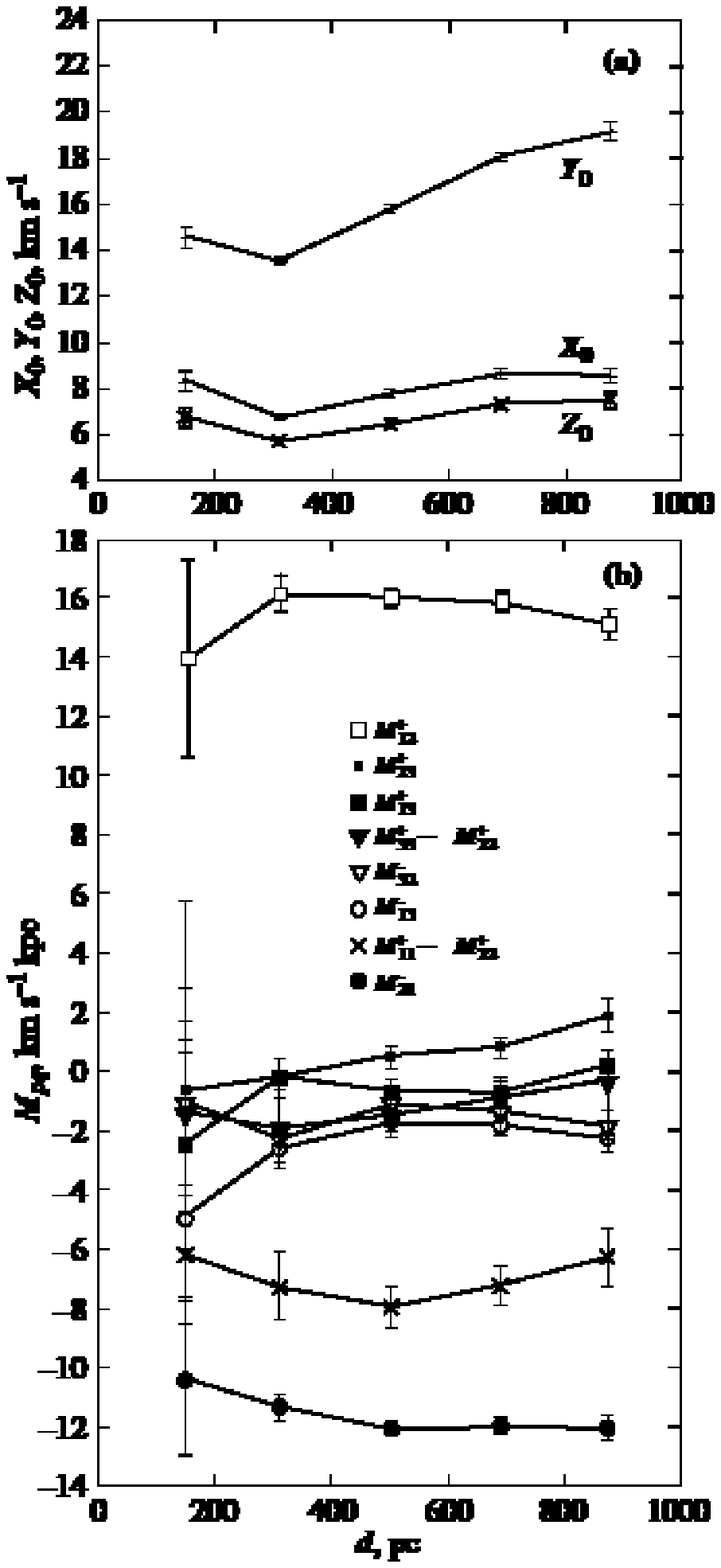}
\end{center}
} Fig.~5. Parameters of the Ogorodnikov-Milne model versus
distance calculated without introducing any model errors.
\end{figure}
%%%%%%%%%%%%%%%%%%%%%%%%%%%%%%%%%%%%%%%%%%%%%%%%%%%%%%%%%%%%%%%

\newpage
%%%%%%%%%%%%%%%%%%%%%%%%%%%%%%%%%%%%%%%% FIG.6:
\begin{figure}[t]
{
\begin{center}
 \includegraphics[width=100mm]{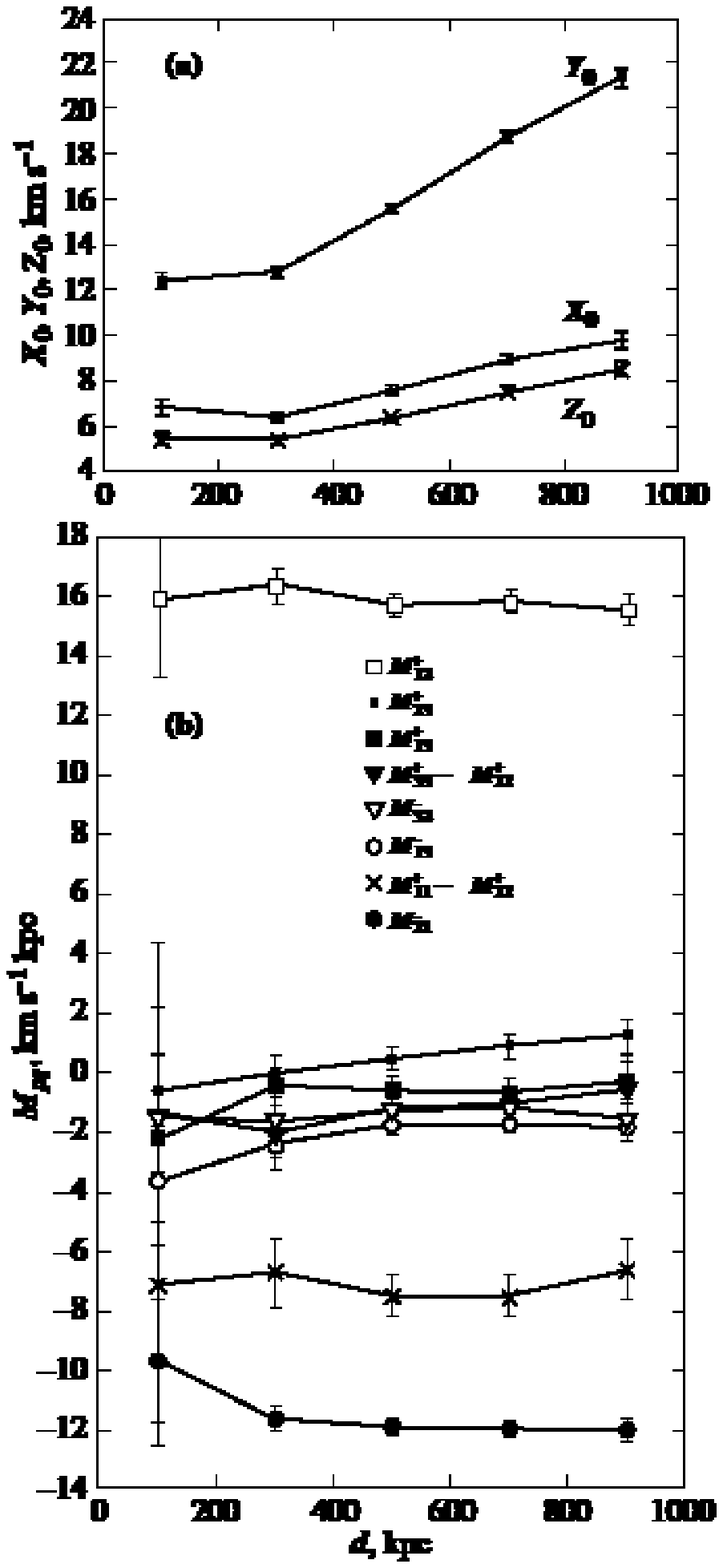}
\end{center}
} Fig.~6. Parameters of the Ogorodnikov-Milne model calculated
with model distances versus distance; the random errors in the
individual distance of a star are 20\%.
\end{figure}
%%%%%%%%%%%%%%%%%%%%%%%%%%%%%%%%%%%%%%%%%%%%%%%%%%%%%%%%%%%%%%%

\newpage
%%%%%%%%%%%%%%%%%%%%%%%%%%%%%%%%%%%%%%%% FIG.7:
\begin{figure}[t]
{
\begin{center}
 \includegraphics[width=100mm]{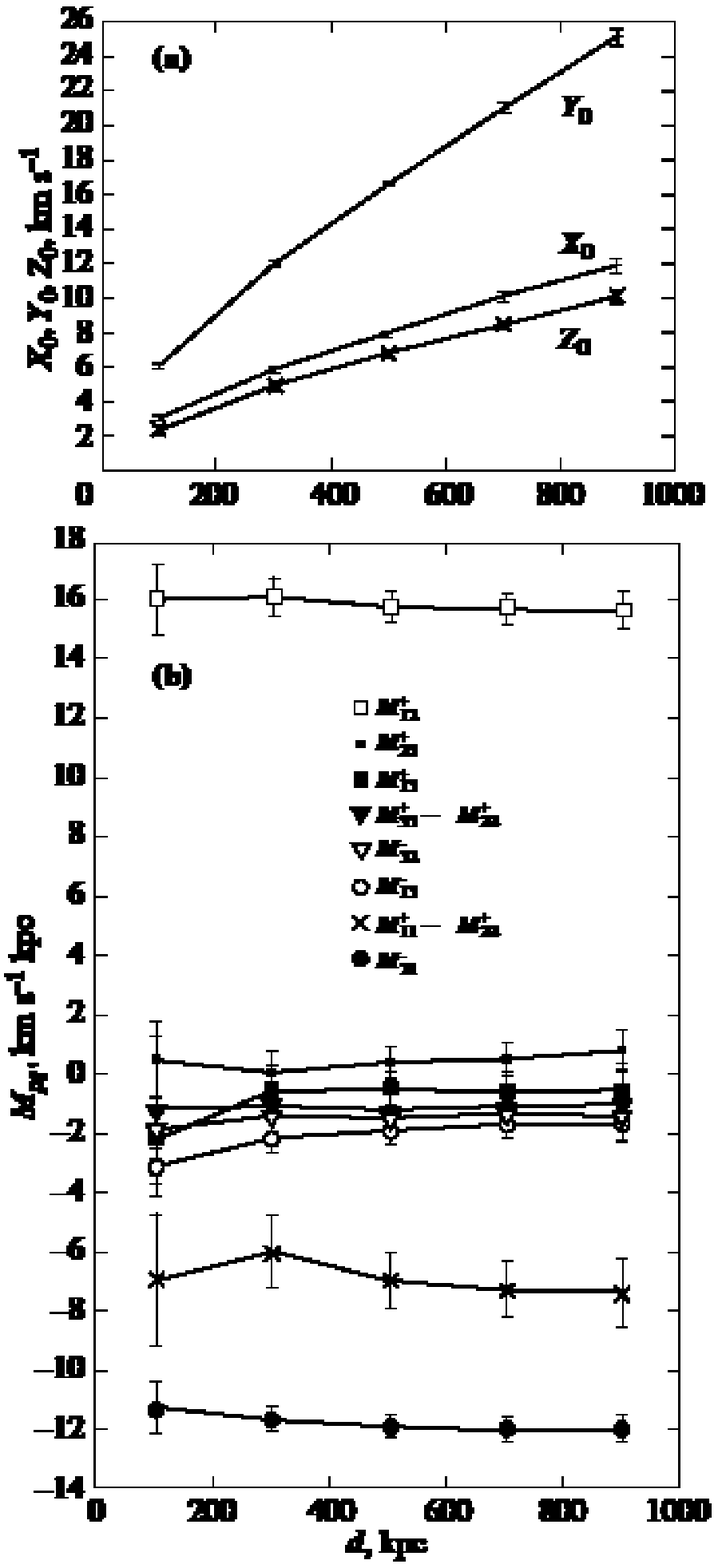}
\end{center}
} Fig.~7. Parameters of the Ogorodnikov-Milne model calculated
with model distances versus distance; the random errors in the
individual distance of a star are 60\%.
\end{figure}
%%%%%%%%%%%%%%%%%%%%%%%%%%%%%%%%%%%%%%%%%%%%%%%%%%%%%%%%%%%%%%%

\end{document}